\documentclass[a4paper,11pt]{article}

\usepackage[margin=1.0in]{geometry}
\usepackage{mathtools}
\usepackage{color}
\usepackage{lipsum}

\definecolor{linkcol}{rgb}{0,0,0} 
\definecolor{citecol}{rgb}{0,0,0} 

\usepackage[bookmarks=true,pdfborder={0 0 0}]{hyperref}
\usepackage[lined,commentsnumbered,ruled,vlined,noend,
linesnumbered, boxruled]{algorithm2e}

\usepackage{graphicx,floatflt}
\usepackage{subfig}
\usepackage{amssymb}
\usepackage{wrapfig}
\usepackage{framed}
\usepackage{caption}

\usepackage[usenames,dvipsnames,svgnames,table]{xcolor}

\newcommand{\blue}{\textcolor{blue}}

\newcommand{\remove}[1]{}
 \newtheorem{theorem}{Theorem}

 \newtheorem{lemma}{Lemma}
 \newtheorem{remark}{Remark}
\newtheorem{invariant}{Invariant}

\newtheorem{defn}{Definition}
\newcommand{\IR}{\mathbb{R}}
  \newenvironment{proof}{\noindent {\bf Proof:\,\ }}{\hfill\mbox{\
  $\Box$}\smallskip}

\newcommand{\FMS}{\textsf{\sc Find-Maximum-Subtree}}
\newcommand{\FMWS}{\textsf{\sc Find-Maximum-Weighted-Subtree}}
\newcommand{\Centroid}{\textsf{\sc Centroid}}
\newcommand{\WeightedMedian}{\textsf{\sc Weighted-Median}}

\newcommand{\EdgeFindOC}{\textsf{\sc Find-Edge-For-Weighted-1-Center}}
\newcommand{\FindSEdge}{\textsf{\sc Find-Split-Edge}}

\title{Facility location problems in the constant work-space read-only
memory model}

\author{Binay K. Bhattacharya\thanks{School of Computing Science, Simon Fraser
University, Canada, {\tt binay@cs.sfu.ca}} \and Minati
De\thanks{The Technion -- Israel Institute of Technology,  Haifa, Israel,
{\tt minati@cs.technion.ac.il}}\and Subhas C.
Nandy\thanks{Indian Statistical Institute,
Kolkata, India, {\tt nandysc@isical.ac.in}}\and Sasanka Roy\thanks{ Chennai
Mathematical Institute,
Chennai, India, {\tt sasanka@cmi.ac.in}}}

\begin{document}
\clearpage
\thispagestyle{empty}
\pagenumbering{roman}
\maketitle
\begin{abstract}
Facility location problems are captivating both from theoretical and
practical point of view. In this paper, we study some fundamental facility
location problems  from the space-efficient perspective. Here the input is
considered to be given in a read-only memory and
only  constant amount of work-space is available during the
computation. This {\em constant-work-space model} is well-motivated for handling
big-data as well as for 
computing in smart portable devices with small amount of extra-space.

First, we propose a strategy to implement prune-and-search in this model.
As a warm up, we illustrate this technique for finding the Euclidean 1-center
constrained on a line for a set of points in $\IR^2$.
This method  works even if the input is given in a sequential access read-only
memory. Using this we show how to compute (i)  the Euclidean 1-center  of a set
of points in $\IR^2$, and (ii) 
the weighted 1-center and weighted 2-center of a tree network. The running time
of all these algorithms are $O(n~poly(\log n))$. 
While the result of (i) gives a positive answer to an open question asked by
Asano, Mulzer, Rote and Wang in 2011, the technique used can be applied to
other problems  which admit solutions by prune-and-search paradigm. For
example,  we can apply the technique to solve two and three dimensional linear
programming  in $O(n~poly(\log n))$ time in this model. To the
best of our knowledge, these are the first sub-quadratic time algorithms for all
the above mentioned problems in the constant-work-space model.
We also present optimal linear
time algorithms for finding the centroid and weighted median of a tree in this
model.

\end{abstract}
\newpage
\cleardoublepage
\pagestyle{plain} 
\pagenumbering{arabic}
\setcounter{page}{1}
\section{Introduction}

The problem of finding the placement of certain number of facilities so that
they can serve all the demands efficiently is  a very important area of
research.  We study   some  fundamental facility location
problems in the memory-constrained environment.
\vspace{-0.08in}
\paragraph{The computational model:} In this paper, we  assume that the input
is given in a read-only memory where modifying the input during the execution is not permissible. This model is referred as {\em read-only
model} in the literature and is studied from as early as 80's~\cite{MunroP78}.
Selection and sorting are well studied in this model~\cite{MunroP78,MunroR96}.

In addition to the read-only model, we assume that 
 only  $O(1)$
extra-space each of  $O(\log n)$ bits is availabe   during the execution. This
is widely known as {\it
log-space}  in the computational complexity class~\cite{AroraB}.
However, we will refer this model as {\it constant-work-space model}
throughout this paper. This model is
well-motivated from the following applications: (i) handling big-data, (ii)
computing in a smart portable devices with small amount of extra-space, (iii) in
a distributed  environment where many procedures access the same data
simultaneously.

In this model, as in \cite{AsanoMW11}, we
assume that a tree $T=(V,E)$ is represented as DCEL (doubly
connected edge list)   in a read-only memory where for a vertex $u\in V$, we can
perform the following queries  in
constant time using constant space:

\vspace{-0.1in}
\begin{itemize}
 \item $Parent(u)$: returns the parent of the vertex $u$ in the tree $T$,
\vspace{-0.05in}
\item $FirstChild(u)$: returns the first child of $u$ in the tree $T$,
\vspace{-0.05in}
\item $NextChild(u,v)$: returns the child of $u$ which is next to $v$ in the
adjacency list of $u$.
\end{itemize}\vspace{-0.1in}
Here we can perform depth-first traversal starting from any vertex in
$O(|V|)$ time.

\vspace{-0.1in}
\paragraph{Definitions and preliminaries:}
 Let $T=(V,E)$ be a tree where $V$ is the set of
vertices (or nodes) and $E$ is the set of edges. The set of points on all the edges of $T$
are also
denoted as $T$. Each vertex $u\in V$ has a weight
$w(u)$ and each edge $e\in E$ has  also a positive length $l(e)$.  For any
vertex $v\in V$, we denote  the degree of $v$ as $d_v$. 
Let  $N(v)$ denote  the set of 
adjacent vertices of $v$. 
 The subtrees  attached to the node $v$ are denoted as
$T_{v'}(v)$, where $v'\in N(v)$.  We denote
$T_{v'}(v^+)=T_{v'}(v) \cup \{\text{the vertex}~ v\} \cup \{\text{the
edge}~ (v,v')\}$.
For any vertex $v \in V$, we denote
$MaxS(v)=\max_{v'\in{N(v)}}|T_{v'}(v)|$, where 
$|T_{v'}(v)|$ denotes the number of vertices in the subtree $T_{v'}(v)$.
The {\it Centroid} of a tree $T=(V,E)$ is a vertex $v^* \in V$ such
that $MaxS(v^*)=\min_{v \in V} MaxS(v)$. This  can be found in $O(n)$ time
using $O(n)$
space~\cite{Ben-MosheBS06,Harary,KH-1}.

For any point $u\in
{T}$, we associate a cost function $SumWD(u)=\sum_{v\in
V}d(u,v)w(v)$,
where $d(u,v)$ is the distance between $u$ and $v$. The {\it weighted median}
of 
${T}$ is defined as a point $x^*$ on the tree ${T}$  such that the associated
cost $SumWD(x^*)$ is minimum over all the points on the edges of the tree 
$T$. Hakimi~\cite{Hakimi65}
showed that there exist a 
weighted median that lies on a vertex of $T$. So, the weighted median is  a
vertex $v$ such that $SumWD(v)=\min_{v'\in V} SumWD(v')$.

For any vertex $v\in {V}$, let $MaxWS(v)=\max_{v'\in N(v)}w(T_{v'}(v))$,
where $w(T_{v'}(v))= \sum_{u\in T_{v'}(v)} w(u)$. The {\it
weighted-centroid} of $T$ is defined as a vertex $v^*$ with
$MaxWS(v^*)=\min_{v\in V}MaxWS(v)$~\cite{KH-2}.   Kariv and Hakimi ~\cite{KH-2}
showed that a vertex $v$ of a tree ${T}$ is  {\it weighted-centroid} if
and only if $v$ is {\it weighted median}.  Based on these facts, they present
an algorithm to find the weighted median of a tree which runs in $O(n)$ time
using $O(n)$ space.

Let $X=\{\alpha_1,\alpha_2,\ldots, \alpha_p\}$  be a set of $p$ points on
the edges of the
tree $T$. For any vertex $v\in V$, by  $d(X,v)$ we mean $\min_{\alpha\in
X}d(\alpha,v)$. The maximum weighted distance from the set
$X$ to tree $T$ is denoted by $S(X,T)$, i.e, 
$S(X,T)=\max_{v\in V} d(X,v)w(v)$.  The {\it weighted $p$-center} of
$T$ is a
$p$ sized subset $X$ of $T$ for which  
$S(X,T)$ is minimum. This problem was originated by Hakimi~\cite{Hakimi65} in
1965 and has a long history in the literature.
 For any constant p, 
an $O(n)$ time  algorithm using  $O(n)$ space is available for this 
problem\cite{Qiaosheng-Shi-Thesis}.

\vspace{-0.11in}
\paragraph{Our main results:}
Prune-and-search is an excellent paradigm to solve different 
optimization problems.
First, we propose a framework to implement prune-and-search in the
constant-work-space model.
As a warm up, we illustrate the technique for finding the Euclidean 1-center
constrained on a line for a set of points in $\IR^2$.
This technique  works even if the input is given in a sequential access
read-only
memory. Using this framework  we show how to compute (i)  the center $c^*$ of
the
minimum enclosing circle for a set of points in $\IR^2$, and (ii)
the weighted 1-center and weighted 2-center of a tree network. The running time
of all these algorithms are $O(n~poly(\log n))$. 
The same framework  can be applied to
other problems  which admits solutions by prune-and-search paradigm. For
example,  we can apply the technique to solve two and three dimensional linear
programming  in $O(n~poly(\log n))$ time in this model. To the
best of our knowledge, these are the first sub-quadratic time algorithms for all
the above mentioned problems in the constant-work-space model.
We also present optimal linear
time algorithms for finding the centroid and weighted median of a tree in this
model.

\vspace{-0.1in}
\paragraph{Related works:}
Constant-work-space model has been studied for a long time and has recently
gained  more attention.  Given an undirected graph testing the existance
of a path between any two vertices~\cite{Reingold05}, planarity
testing~\cite{AllenderM04}, etc. are some of the important problems for which 
outstanding results on constant-work-space algorithms are available. Selection
and sorting are extensively studied in the read-only model~\cite{MunroR96}.
Specially, we want to mention the pioneering work by Munro and
Paterson~\cite{MunroP78}, where they proposed 
$O(n\log^3n)$ time constant-work-space algorithm for the selection considering
that the input is given in a sequential access read-only memory.

 Our work was inspired by an open question from Asano et al.~\cite{AsanoMRW11}
where they presented several constant-work-space algorithms  for geometric
problems like geodesic shortest path in a simple polygon, Euclidean minimum
spanning tree, and they asked for any sub-quadratic time algorithm for minimum
enclosing circle in the constant-work-space model.  De et al.~\cite{DeNR12b}
presented a sub-quadratic time algorithm for the problem using $\Omega(\log n)$ 
extra-space in the read-only model.  An $1.22$ approximation algorithm for
minimum enclosing ball for a set of points in $\IR^d$ using $O(d)$ space is
known~\cite{AgarwalS10,ChanP11} in the streaming model where only one pass is
allowed in  the sequential access read-only input. For fixed dimensional linear
programming, Chan and Chen~\cite{ChanC07} presented a randomized algorithm in
expected $O(n)$ time using $O(\log n)$ extra-space in the read-only model. We
refer \cite{AsanoBBKMRS13} for other related recent works in the literature. 

\vspace{-0.1in}
\section{Prune-and-search using constant work-space}\vspace{-0.1in}
A general scheme for implementing prune-and-search when the input is given in a
read-only array is presented in \cite{De-Minati-Thesis,DeNR12b} using
$\Omega(\log n)$ extra space. 
Prune-and-search is an iterative  algorithmic paradigm. Initially, all
the input elements are considered {\it valid} and after each iteration a
fraction of the valid elements are identified whose deletion will not impact on
the optimum result. So, these elements are pruned, and the
 process is repeated with the reduced set of valid elements until  the
desired optimum result is obtained or the number of valid elements is a small
constant. For the later case, a brute-force search is applied for
obtaining the desired result. 

In constant-work-space model, after each iteration, we have to distinguish 
the valid and pruned elements correctly using only $O(1)$ space. Here, we
demonstrate that in some
special cases, where 
the combinatorial complexity of the 
feasible region is
$O(1)$ after each iteration, the prune and search can be implemented using
$O(1)$ extra space. As a warm up, we describe this using the
prune-and-search algorithm for finding the Euclidean 1-center constrained on a
line $L$ for a set of
points in $\IR^2$ ~\cite{Megiddo83a}.

\vspace{-0.1in}
\subsection{Constrained Euclidean 1-center}\vspace{-0.09in}
\label{CMEC}
A set of $n$ points $P$ in $\IR^2$ is given in a read-only array $P$ and a
vertical line $L$ is given as a query. The objective is to find a point $x^*$ on
the line $L$ such that the maximum distance of $x^*$ from the points of $P$ is
minimized
over all possible points on $L$.
\vspace{-0.1in}
\paragraph{Megiddo's Algorithm~\cite{Megiddo83a}:} Initially, all the input
points are considered
as  valid. In an
iteration, if $n'$ is the number of valid
elements, then $\frac{n'}{2}$ disjoint pairs are formed. Each of these pairs
contributes a perpendicular bisector that intersects the line $L$. Considering these $\frac{n'}{2}$ intersection points on the line $L$, 
the algorithm finds the median intersection point $m$ among them.  Then it makes
the  following query:

\noindent
{\it Query($m$):}- decide whether $x^*=m$ or $x^*$ lies
above or below  of $m$ on $L$. We compute the
farthest point(s) from $m$ among all the valid points. If there exists
two farthest points  above and below $m$ respectively,
then  $x^*=m$; otherwise, if all the
farthest points are in one side, say above (resp. below) of $m$, then 
$x^*$ lies above (resp. below) of $m$ on $L$.

\noindent
In the former case, the algorithm stops, and in the later case, from each pair
whose corresponding perpendicular bisector
intersects the line
$L$ below (resp. above) $m$, one element is pruned.
Thus, in a single
iteration,
$\frac{n'}{4}$ points can be pruned. 
The next iteration is
executed on the remaining valid points unless very few (say, 3 or 4) elements
remain as valid, in which case brute-force search is applied to find $x^*$.

\begin{wrapfigure}{r}{0.4\textwidth}
\vspace{-0.5in}
\begin{center}
    \includegraphics[width=0.39\textwidth]{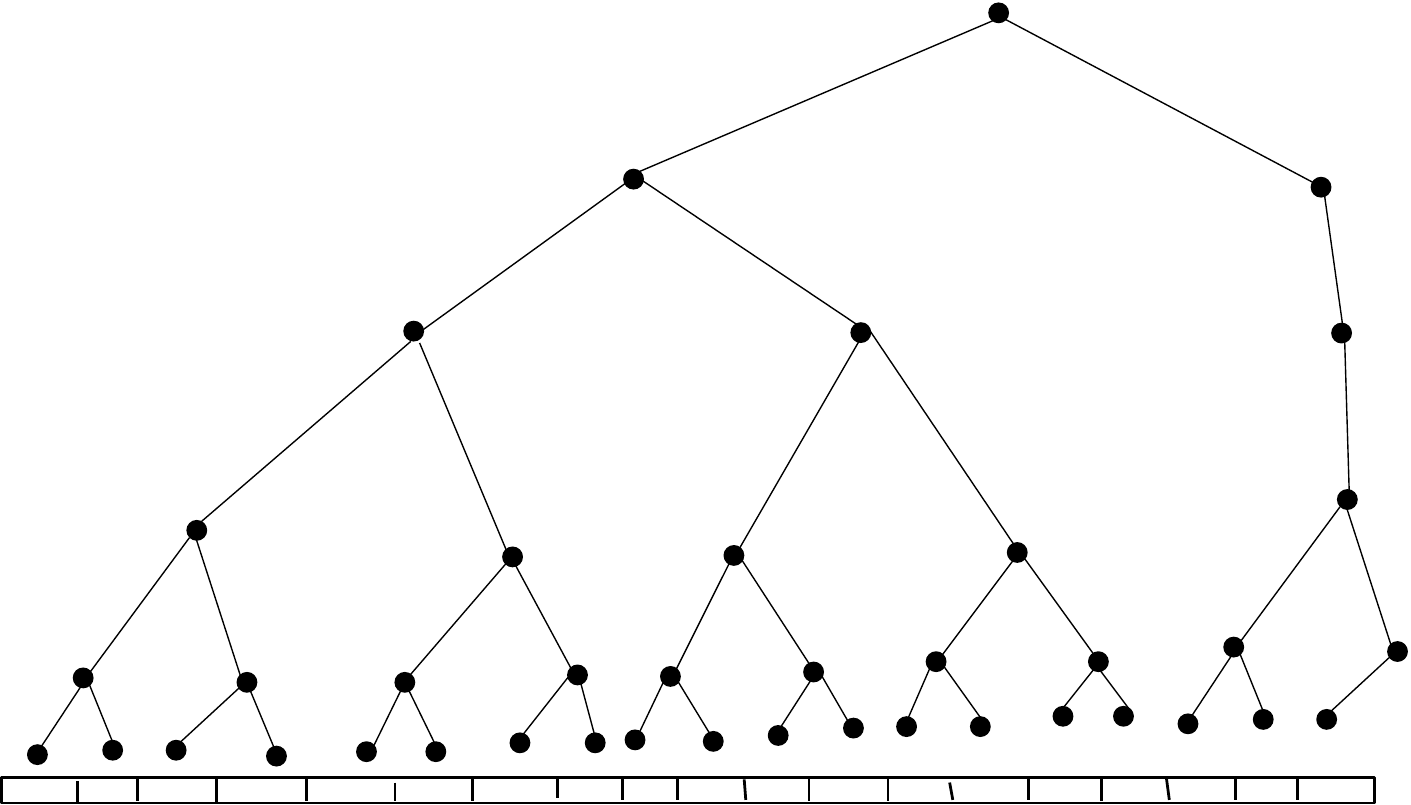}
  \end{center}
  
    \caption{The input array and the corresponding virtual pairing tree}
    \vspace{-0.2in}
  \label{fig1}
\end{wrapfigure}

\vspace{-0.1in}
\paragraph{Overview of our pairing scheme:}

We are going to describe a pairing strategy for the prune-and-search algorithm
consisting of $O(\log n)$ phase and each phase consists of at most $O(\log n)$ 
iterations.  After each iteration, it will remember a {\em feasible region}
$U=(a,b)$ on the line $L$  such that the point $x^*$ lies
in the
region $(a,b)$. Initially, $a=-\infty,b=+\infty$, and they are updated after
each iteration. This information will help to distinguish the valid and pruned
elements.
Consider the {\it virtual pairing tree}, ${\cal T}$ (See Figure~\ref{fig1}),  
which
is a
binary tree of depth $d=\lceil \log n\rceil$  and  leaves are the input
points stored in the read-only array $P$. The subtree rooted at any node $t$ is
denoted
as $T_t$, and 
the leaves of the tree $T_t$ are denoted as $\beta({T_t})$.
The nodes in the
$k$-th ($0 < k < d$) level (assuming the leaves are at
$1$-st level) represent all the valid points at the
beginning of the $k$-th phase of the algorithm. So, the nodes at $k$-th level
are actually a subset of valid nodes of $(k-1)$-th level.
Any  node $t$ at $k$-th level is   the only one among $\beta({T_t})$
 which is valid after $(k-1)$-th phase of the algorithm. 
The algorithm stops
pruning when  very few (say,     3  or   4)  elements  are  valid or already
$x^*$ is found. This virtual pairing tree demonstrates how the pairing
is done in each phase of the algorithm.


%

%

Let $U=(a,b)$ be the feasible region for the constrained Euclidean 1-center
$x^*$ on the line $L$. We define a {\em dominance} relation as follows. 

\vspace{-0.07in}
\begin{defn}\label{def1}
For a pair of points $p,q \in P$,
  $p$ is said to {\em dominate} $q$ with respect to a
feasible region $U$, if their perpendicular bisector $b(p,q)$ does not
intersect the feasible region $U$, and both $q$ and $U$ lie on the same
side of
$b(p,q)$.
\end{defn}
\vspace{-0.07in}
It is easy to show from Definition \ref{def1} that $p$ dominates $q$ with
respect to a feasible region $U$ if and only if  from any point $x\in U$,
$\widehat{d}(p,x)>\widehat{d}(q,x)$, where $\widehat{d}(p,x)$ (resp.
$\widehat{d}(q,x)$) is Euclidean distance between $p$ (resp. $q$) and $x$.  

\vspace{-0.07in}
\begin{lemma}
 \label{L0}
 If $p$ dominates $q$ and $q$ dominates $r$ with respect to a feasible region
$U$, then $p$ dominates $r$ with respect to the feasible region
$U$.
\end{lemma}
\vspace{-0.07in}
\begin{proof}
Let $x$ be an arbitrary point in $U$. Since $p$ dominates $q$, 
$\widehat{d}(p,x)> \widehat{d}(q,x)$. 
 Since $q$ dominates $r$, $\widehat{d}(q,x) > \widehat{d}(r,x)$. Thus
$\widehat{d}(p,x) > \widehat{d}(r,x)$. 
\end{proof}

\vspace{-0.07in}

\vspace{-0.07in}
\paragraph{First phase} The feasible region $U=(a,b)$ is initialized as
$(-\infty,\infty)$. Now, all the points in $P$ are valid. We form
pairs, $Pair_i^1=(P[2i-1],P[2i])$, $i=1,2,\ldots, \lfloor\frac{n}{2}\rfloor$.
A pair $Pair_i^1$, $i \in \{1,2 \ldots, \lfloor
\frac{n}{2} \rfloor \}$ is considered to be a {\it valid pair} with respect to
$(a,b)$
if the corresponding perpendicular bisector intersects $(a,b)$ on the  line
$L$. 

In an iteration, we consider only the valid pairs with respect to
$(a,b)$ (in the first  iteration, all the pairs are valid). 
Considering the intersection points  of the perpendicular bisectors of
these valid pairs with $L$, we compute 
 the
median intersection point $m$ on $L$.
Then we perform {\it Query($m$)} by
inspecting all the points of $P$ as  in Megiddo's algorithm.
Depending on the answer of the query, either $x^*$ is found, or $U$ is updated
by assigning $a$ or $b$  with $m$. In the former case, the algorithm stops and
in the later case, from each of the pairs whose corresponding perpendicular
bisector intersects $L$ outside the revised $(a,b)$, one element is pruned. So,
after
this iteration, one element
each  from at least $\frac{1}{4}$-th of the valid pairs is 
pruned. The
algorithm executes next iteration with the remaining valid pairs. The process continues
 until from each pair 
$Pair_i^1$, $i=1,2,\ldots, \lfloor\frac{n}{2}\rfloor$, one
element is pruned.    Since, after each iteration, one
element
from at least $\frac{1}{4}$-th of the valid pairs is 
pruned,  this phase executes at most $O(\log n)$ iterations. Finally, after completion of this phase, we can discard $\lfloor\frac{n}{2}\rfloor$ points, i.e, one point from each of the pair $Pair_i^1$, $i=1,2,\ldots, \lfloor\frac{n}{2}\rfloor$.


\vspace{-0.1in}

\paragraph{$k$-th phase} 
At the beginning of the $k$-th phase, let $U=(a,b)$ and  we know that only one
element is valid  (i.e. dominant) from each block of consecutive $2^{k-1}$
elements, namely
$B_i^k=\{P[i.2^{k-1}+1],P[i.2^{k-1}+2], \ldots, P[(i+1).2^{k-1}]\}$,
$i=1,2,\ldots,
\lfloor
\frac{n}{2^{k-1}}\rfloor$. For the last block
$B_{\lceil\frac{n}{2^{k-1}}\rceil}^k$, the
members are $\{P[i.2^{k-1}+1],P[i.2^{k-1}+2], \ldots P[n]\}$.
We denote the only valid element of a block $B_i^k$ as $valid(B_i^k)$. 
Now, the most
important task is to
recognize the $valid(B_i^k)$ for all  $i= 1,2,\ldots,
\lceil
\frac{n}{2^{k-1}}\rceil$.
In this regard, we have the
following:

%
%
%

\vspace{-0.1in}
\begin{lemma}
\label{L2}
 The $valid(B_i^k)$    can be identified in $O(|B_i^k|)$ time
using $O(1)$ extra-space, where $i=
1,2,\ldots,
\lceil
\frac{n}{2^{k-1}}\rceil$ and $0<k<d$.
 
\end{lemma}\vspace{-0.1in}
\begin{proof}
The transitivity of the dominance relation (see Lemma~\ref{L0}) and
the pairing strategy guarantee that $valid(B_i^k)$ dominates all other elements
of the block $B_i^k$.
For the block $B_i^k$, we initialize two variables  $candidate=i.2^{k-1}+1$ and $pointer=i.2^{k-1}+2$.

In the first step, we pair up $(P[candidate],P[pointer])$ and observe their perpendicular bisector. Here either of the two situation occurs.
(i) If the perpendicular bisector of the pair intersects the feasible region
$U$, then none of these two points is $valid(B_i^k)$\footnote{If any one of
$P[candidate]$ or $P[pointer]$ is $valid(B_i^k)$, then their perpendicular
bisector would intersect outside U.}.
We update $candidate=pointer+1$ and $pointer=pointer+2$. 
(ii) Otherwise, one of the points of $(P[candidate],P[pointer])$ dominates the other. We update the variable $candidate$ with  the index of the dominating
 one  and $pointer=pointer+1$.

We repeat the next step with new pair until the variable $pointer$ reaches the last element of the block $B_i^k$.
At the end, we obtain  $valid(B_i^k)=P[candidate]$. 
Thus, the lemma follows. 
\end{proof}
\vspace{-0.05in}
Thus, in this phase, we can correctly enumerate all the valid elements in
$O(n)$ time using $O(1)$ space.

As in the first phase, we construct pairs
$Pair_i^k=(valid(B_{2i-1}^k),valid(B_{2i}^k))$, for $i= 1,2,\ldots,
\lceil \frac{n}{2^{k-1}}\rceil$, and consider a pair $Pair_i^k$ to be a {\it
valid pair} with respect to $(a,b)$ if the corresponding perpendicular
bisector intersects $(a,b)$ on $L$. Here also  we need at most 
$O(\log( \lceil \frac{n}{2^{k-1}}\rceil))$ iterations to discard one element 
from each $Pair_i^k$. Needless to say, during this process $x^*$ may also be
found. 
Thus, after this phase from each of the block $B_i^{k+1}$, $i= 0,1, \ldots
\lceil
\frac{n}{2^{k}}\rceil$, only one element survives.


As the depth of the virtual pairing tree is $O(\log n)$, so there are at most
$O(\log n)$ phases. Each phase needs at most $O(\log n)$  iterations, each of
which needs $O(n +M)$ time. Here $M$ is the time needed to compute the median of
$n$ elements in the read-only memory when $O(1)$ space is
provided~\cite{MunroP78,MunroR96,RamanR99}. Thus  we have the
following theorem.
\vspace{-0.1in}
\begin{theorem}\label{th1}
 Given a set of $n$ points in $\IR^2$ in a read-only memory and a line $L$, 
 the Euclidean 1-center constrained on a line $L$ can be found
in $O((n+M)\log^2 n)$ time using $O(1)$ extra-space, where $M$ is the time
needed to compute the median of
$n$ elements given in a read-only memory using $O(1)$ extra-space.
\end{theorem}
\vspace{-0.1in}

%
%
%
%

\vspace{-0.1in}
\begin{remark}
Note that our pairing strategy will work even if the input is given in sequential
access read-only memory.
 Here the median finding algorithm is appropriately chosen 
for the sequential access read-only
memory \cite{MunroP78}. For
detailed literature on selection, we refer~\cite{Chan10}.
\end{remark}

\vspace{-0.2in}
\section{Euclidean 1-Center}
\vspace{-0.1in}
\paragraph{Problem Statement}
The {\it Euclidean 1-center} of a set $P$ of $n$ points in $\IR^2$ is a
point
$c^*\in \IR^2$ for which the maximum distance from any point in $P$ is
minimized. The point $c^*$ is actually the center of the {\it minimum enclosing
circle} of $P$. Here,   
we assume that the input is  given in a
read-only memory and only constant amount of work-space   is
available for the computation.

\vspace{-0.1in}
\paragraph{Megiddo's Algorithm~\cite{Megiddo83a}:}
This is a prune-and-search algorithm that uses the following sub-routine.
{\it Decide-on-a-Line($L$)}: Given a set of points $P$ in $\IR^2$
and a query line $L$, decide in which side of $L$ the Euclidean 1-center $c^*$ for the points in $P$
lies.

In Megiddo's algorithm, initially all the input points are considered to be valid. In
an iteration, if $n'$ is the number of valid points, then
$\lfloor\frac{n'}{2}\rfloor$ disjoint pairs are formed. Each pair 
contributes a perpendicular bisector. Let us denote this set of
perpendicular bisectors as ${\cal P_B}$. Compute the median slope $S_m$ of these bisectors. 
A perpendicular bisector $b_i\in {\cal P_B}$ with slope less than $S_m$  is
paired with a perpendicular bisector $b_i\in {\cal P_B}$ having slope greater
than or equal to $S_m$. In this way, $\lfloor\frac{n'}{4}\rfloor$ disjoint
pairs of bisectors are formed. Each pair of bisectors contribute an intersection
points. So there are $\lfloor\frac{n'}{4}\rfloor$ intersection points. Let us
denote this set of intersection points as $I$.  The intersection
point $t\in I$
with median $x$-coordinate (with respect to rotated coordinate system by an
angle $S_m$) is identified. Now, the subroutine {\it
Decide-on-a-Line} is evoked for the line $L_1$ passing through $t$ with slope
$S_m$ to decide in which side of $L_1$ the point $c^*$ lies. Next, consider the intersection
points of $I$ which lies to the  side of $L$ opposite to $c^*$, and find the
intersection point $t'$ having median $y$-coordinate value (with respect to
rotated coordinate system by an angle $S_m$). 
Let $L_2$ be the line perpendicular to $L_1$ and passing through $t'$. We evoke the 
subroutine {\it Decide-on-a-Line} for the line $L_2$. 
Thus, a quadrant $Q$ is defined by the two lines $L_1$ and $L_2$ which
contains $c^*$. The choice of the lines $L_1$ and $L_2$
guarantees that $\frac{n'}{16}$ perpendicular bisectors from ${\cal P_B}$ will
not intersect the quadrant $Q$. This allows us to prune a  point corresponding to 
each of those perpendicular bisectors. As a result, after each iteration at
least $\frac{n'}{16}$ of the valid points are pruned. The iteration is repeated
for the rest of the valid points until the number of valid points become very
small (say 15), or already $c^*$ is found. In the former case, brute-force is
applied to compute the point $c^*$. 

\vspace{-0.1in}
\paragraph{Our implementation of the algorithm in constant work-space model:}
First, we show that  {\it Decide-on-a-Line($L$)} can be answered using
$O(1)$ extra-space.
\vspace{-0.1in}
\begin{lemma}\label{LD}
For a set of $n$ points $P$ in $\IR^2$, {\em Decide-on-a-Line($L$)}  can be 
computed in $O((n+M)\log^2 n)$ time using $O(1)$ extra-space, where $M$   is the
time needed to compute the median of $n$ elements given in a read-only memory
using $O(1)$ extra-space.
\end{lemma}\vspace{-0.1in}
\begin{proof}
By Theorem~\ref{th1}, we can compute the constrained Euclidean 1-center $x^*$ 
on the line $L$ in $O((n+M)\log^2 n)$ time using $O(1)$ extra-space. Now, in a
single scan over all the points in $P$, we can identify the farthest point(s)
from $x^*$. Let $F$ be the set of points that are farthest from $x^*$. 
\begin{description}
\item[Step 1:]By scanning the whole array, we can decide whether all the points
in $F$ are in one side of $L$. If the test is positive, then $c^*$ will be in
the same side of $L$; otherwise we go to the next step.
\vspace{-0.05in}
\item[Step 2:] Now, the points of $F$ are in both side of the
line $L$. If the convex hull defined by $F$ contains $x^*$, then 
$c^*=x^*$. 
For this, we do not have to construct the convex hull
explicitly. Let $t_1$ and  $b_1$ (resp. $t_2$ and $b_2$) be the two points of $F$
in one side of $L$ whose projections on $L$ are the farthest apart.  Consider
two lines joining $t_1,t_2$  and $b_1,b_2$ and observe  their intersections 
with the line $L$.  
The convex hull of $F$ contains $x^*$ if and only if $x^*$ is in between these
two intersection points because the points in $F$ are in a circle whose center
is $x^*$. In the positive case, $c^*=x^*$. Otherwise, we go to the next step.
\vspace{-0.05in}
\item[Step 3:]Now, the midpoint of the line joining the farthest
pair of points in $F$ will determine the side of $L$ in which $c^*$ lies. In
this case either $(t_1,t_2)$ or $(b_1,b_2)$ are the farthest pair of points in
$F$.  
\end{description}
\vspace{-0.1in}
Thus the lemma follows.
\end{proof}

We implement Megiddo's algorithm in a similar way as described in Section~\ref{CMEC}.
 Note that in our scheme, we need to remember a {\em feasible
region} $U$ for $c^*$ of constant combinatorial complexity after each iteration. Here, after each iteration, 
we get a quadrant (defined by a pair of mutually perpendicular lines $L_1$
and $L_2$) that contains the $c^*$. But considering all the iterations, the
intersection of all these quadrants has combinatorial complexity $O(\log n)$.
So, the straight-forward implementation will not lead to an algorithm in
constant-work-space model. To overcome this, we apply the following simple
trick. After each iteration we will remember a feasible region $U$ as a
triangle. After the first iteration of the algorithm we have a quadrant $Q$ in
which $c^*$ lies. We obtain a triangle $T\subset Q$ containing $c^*$  using
the following lemmas.
%
\vspace{-0.1in}
\begin{lemma}\label{mec:l2}
 If we know a quadrant $Q$ in which $c^*$ lies, then  we can obtain a
triangle $T\subseteq Q$ containing $c^*$ by evoking  the
subroutine {\it Decide-on-a-Line($L$)}  once more.

\end{lemma}\vspace{-0.1in}
\begin{proof} 
  We scan the points in $P$ to find the axis-parallel 
rectangle $R$ containing all the points in $P$. Observe that $R$ contains
$c^*$. The polygon $R'=R\cap Q$ has at most four side. If $R'$ is a triangle,
then $T=R'$. Otherwise,   
  by evoking  the
subroutine {\it Decide-on-a-Line($L$)} on any diagonal of the  quadrangle $R'$,
we can decide  a triangle $T\subset R'$ containing $c^*$.
\end{proof}

\remove{At the beginning of an iteration, we have a feasible triangular region
$T$, and after that iteration we get a new  quadrant $Q$
containing $c^*$. We can again obtain a triangle
$T'\subseteq  T\cap Q$  containing $c^*$ by following lemma. }

\vspace{-0.1in}
\begin{lemma}\label{mec:l3}
 Let $T$  be a triangle and $Q$ be a quadrant both of which
contain the Euclidean 1-center $c^*$. We can obtain
another triangle $T' \subseteq T\cap Q$  containing $c^*$  by evoking  the
subroutine {\it Decide-on-a-Line($L$)} at most twice.
\end{lemma}\vspace{-0.1in}
\begin{proof}
Note that the $R=T\cap Q$ is a polygon with at most five sides. So, we can
triangulate the polygon $R$ using at most two diagonals. By evoking {\it
Decide-on-a-Line($L$)} on each of these diagonals, we can decide the triangle
$T'$ containing $c^*$.
\end{proof}

%

Observe that any perpendicular bisector which does not intersect $Q$ and $T$ also does
not intersect $T'$. Thus, we remember 
a triangular feasible  region $U$ for $c^*$ using $O(1)$ extra-space.
Now, we  apply our constant-work-space pairing strategy to this
modified algorithm.
Here, again we have $O(\log n)$ phase each consisting of $O(\log n)$ iterations. In the beginning of the $k$-th phase, we know only one element is valid from each block $B_i^k$ of consecutive $2^{k-1}$ elements.  
Similar to Definition~\ref{def1}, here also we can define {\em dominance}
relation with respect to the feasible region $U$ and it is easy to prove that Lemma~\ref{L0} and \ref{L2} hold. 
We construct pairs $Pair_i^k=(valid(B_{2i-1}^k),valid(B_{2i}^k))$, for $i=1,2,\ldots,\lceil\frac{n}{2^{k-1}}\rceil$. 
A pair $Pair_i^k$ is considered a {\em valid pair} with respect to the feasible region $U$, if the perpendicular  bisector of that pair intersects $U$.
In an iteration,  by making at most four calls to the 
subroutine  {\it
Decide-on-a-Line($L$)}, we update our feasible region $U$ which guarantees that one point each from  at least $\frac{1}{16}$-fraction of the valid pairs are pruned with respect to $U$.
As each iteration takes $O((n+M) \log^2 n)$ time and $O(1)$ extra-space (by Lemma~\ref{LD}, \ref{mec:l2} and \ref{mec:l3}) and there are at
most $O(\log^2 n)$ iterations,  the running time of this algorithm is $O((n+M)\log^4 n)$, 
where $M$ is the time required to compute the median of a set of $n$
elements in the constant work-space model. Thus we have the  following result.

\vspace{-0.1in}
\begin{theorem}
 Given a set of $n$ points in $\IR^2$, we can compute the Euclidean
1-center in $O((n+M) \log^4 n)$ in the constant-work-space model,
where $M$ is the time required to compute the median of $n$ elements in the 
constant work-space model.

\end{theorem}
\vspace{-0.1in}

\section{Centroid of a tree}
\label{centroid}
The quadratic time  algorithm for finding the
centroid of a tree $T=(V,E)$ in the constant-work-space model is quite obvious.
For each vertex
$v \in V$, compute $MaxS(v)=\max_{v'\in N(v)}|T_{v'}(v)|$ 
by inspecting all its neighbors' subtree, and finally report the 
centroid of $T$ which is a vertex  with minimum $MaxS(v)$ value.

\begin{wrapfigure}{r}{0.64\textwidth}
\vspace{-0.4in}
\begin{center}
    \includegraphics[width=0.63\textwidth]{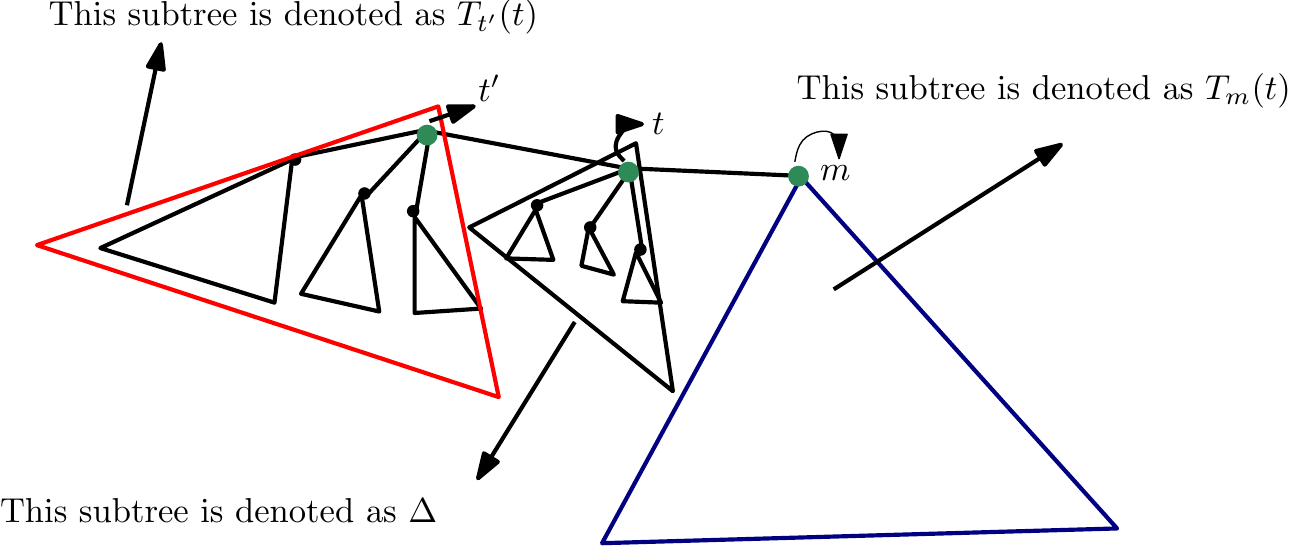}
  \end{center}
    \vspace{-0.1in}
    \caption{Illustration for the definitions}
    \vspace{-0.15in}
  \label{fig2}
 
\end{wrapfigure}

In this section, we present a linear time algorithm
for finding the centroid of a tree $T$ using only constant amount of
extra-space. This algorithm is similar to the $O(n)$ time and $O(n)$
space algorithm given in~\cite{KH-1} for the same problem. Here $n=|V|$. It is based on the fact that
a vertex $v^*\in V$ is the  centroid if
and only if $MaxS(v^*)\leq\lfloor\frac{n}{2}\rfloor$ (see~\cite{Harary}).


Our algorithm starts from an arbitrary vertex $t$ (say the
$t=root(T)$) and finds it's  adjacent vertex $m$ such that
$|T_m(t)|=MaxS(t)=\max_{v'\in N(v)}|T_{v'}(t)|$. If $|T_m(t)|\leq
\lfloor\frac{n}{2}\rfloor$, then $t$ is the centroid; otherwise $t$
can not be
the centroid and the centroid
must be in the subtree $T_m(t)$. In the later
case, we will continue to search in the subtree $T_m(t)$ ignoring $T\setminus
T_m(t)$.   The pseudo-code of our algorithm is given in Algorithm
\ref{Alg-Centroid}. 
Here, we  use the variables $t$, $t'$ and $Size$, maintaining the following
invariant. 
%
\vspace{-0.1in}
\begin{invariant}\label{inv1}
 
\begin{itemize}
 \item Initially,  $t=root(T)$, $t'=\emptyset$ and $Size=0$.

\item If $t'\neq \emptyset$, then 
$t$ and $t'$ are adjacent vertices of $T$ and $Size=|T_{t'}(t)|$
\end{itemize}
\end{invariant}
\vspace{-0.1in}

%
%
%

\begin{wrapfigure}{R}{0.5\textwidth}
    \vspace{-0.2in}
    \begin{minipage}{0.5\textwidth}
\begin{algorithm}[H]
\scriptsize
\SetAlgoLined
\caption{\Centroid($T$)}
\label{Alg-Centroid}
\KwIn{A tree $T$ is given in a read-only memory}
\KwOut{Report the centroid of the given tree}
\SetKwRepeat{Repeat}{do}{while}
$t=root(T)$; $t'=\emptyset$; $Size=0$\;

\Repeat{$TMsize \leq\lfloor
\frac{n}{2}\rfloor$}
{
$(m,\delta, TMsize)=\FMS(t,t',Size)$\;
\If{$TMsize > \lfloor \frac{n}{2}\rfloor$}
{$t'=t$; $t= m$; $Size= Size +\delta$ \;}

}
Return $t$\;
\normalfont
\end{algorithm}
\end{minipage}
    \vspace{-0.2in}
  \end{wrapfigure}

At each iteration of the do-while loop, the  algorithm evokes the procedure
\FMS($t,t',Size$) which  returns three parameters $m$, $\delta$ and $TMsize$.
Here $m$ is the adjacent vertex of $t$ such that
$|T_m(t)|=MaxS(t)=\max_{v'\in
N(v)}|T_{v'}(t)|$. Here $TMsize=|T_{m}(t)|$ and $\delta$ is the number of
vertices in the subtree  $\Delta= T\setminus\{T_{t'}(t)\cup
T_{m}(t) \}$ (see Figure~\ref{fig2}).  Now, depending on the value of $TMsize$,
following  two cases arise.
\begin{itemize}
 \item If $TMsize \leq\lfloor \frac{n}{2}\rfloor$, then $t$ is the centroid. In
this case, the algorithm  stops execution after
reporting  $t$.
\item Otherwise (i.e $TMsize > \lfloor \frac{n}{2}\rfloor$),  $t$ is not the
centroid. Here  $t'$ is updated to $t$; $t$ is updated to $m$ and $Size$ is
incremented by $\delta$. Then
 it repeats the iteration of the do-while loop.
\end{itemize}\vspace{-0.09in}

Note that, as $TMsize+ Size=|T_{m}(t)|+|T_{t'}(t)|$ and the value of  $Size$
monotonically increases after each iteration, the loop will definitely
terminate. The correctness of the algorithm follows from
the fact that a centroid can not have a subtree of size greater than
$\lfloor \frac{n}{2}\rfloor$~\cite{Harary}.
%


\vspace{-0.1in}
\begin{lemma}\label{l4}
 The procedure \FMS($t,t', Size$), which returns three parameters $m$, $\delta$
and $TMsize$,  can be implemented in $O(2\delta)$  time using $O(1)$
extra-space. Here  $m$
is vertex adjacent to $t$ such that $|T_m(t)|=\max_{v'\in
N(v}|T_{v'}(t)|$, $TMsize=|T_{m}(t)|$ and $\delta$ is the number of vertices in
the subtree $\Delta$, where $\Delta= T\setminus\{T_{t'}(t)\cup
T_{m}(t) \}$ (see Figure~\ref{fig2}).
\end{lemma}\vspace{-0.18in}
\begin{proof}
We  implement the procedure \FMS($t,t', Size$)
by a similar way  as Asano et al.~\cite{AsanoMW11} did for their
{\it FindFeasibleSubtree}. 
Note that using the three routines, namely
$Parent(v')$, $FirstChild(v')$, 
$NextChild(t,v')$,  we can compute the number of vertices in the
subtree $T_{v'}(t)$ by a depth-first traversal for any vertex $v'\in V$ . It
takes $O(|T_{v'}(t)|)$
 time and  $O(1)$ extra-space.

We maintain a pointer variable $m$ and two integer variables $TMsize$ and $ENC$.
They are initialized as $m=t'$, $TMsize=Size$ and $ENC=0$, respectively. At any
moment during the execution of the procedure, $m$, $TMsize$ and $ENC$ signify,
so far obtained, the subtree with maximum size, the size of the maximum sized
subtree and the number of vertices traversed, respectively.

As $Size$ signifies $|T_{t'}(t)|$
(by Invariant~\ref{inv1}), we already know the size of the subtree $T_{t'}(t)$.
So, we do not need to  perform a depth-first-traversal in the subtree
$T_{t'}(t)$. We start computing the number of vertices for two
subtrees in parallel 
using two pointers $\pi_1$ and $\pi_2$ (in sequential machine one move of 
$\pi_1$ is followed by one move in $\pi_2$, and vice versa). While traversing 
a subtree by $\pi_i$, its root is stored in $\phi_i$, $i=1,2$. For each
$\pi_i$, a variable $\chi_i$ is maintained that stores the number of vertices
encountered  by $\pi_i$. If one of $\pi_1$ and $\pi_2$ completes its 
task in a subtree, then it start traversal in the next
unprocessed subtree. After each step, $ENC$, $\chi_1,\chi_2$ are updated. The
variables $m$ and  $TMsize$  are updated accordingly. The process terminates
when one of $\pi_i$ (say $\pi_1$) finds that there is no more subtree to
process. Thus the remaining subtree $T_{\phi_2}(t)$ of the other pointer
($\pi_2$) is not traversed completely, but we can compute the number of vertices
in that subtree as $n-Size-ENC+\chi_2$, and update $m$ and $TMsize$, if needed.
We compute the number of vertices in the subtree $\Delta$
as $\delta=(n-Size-TMsize)$.

 Thus, when the process
stops, both $\pi_1$ and $\pi_2$ 
traversed equal number of elements. As  all the
subtrees in $\Delta$ can be processed in at most $2\delta$ steps, so the time
needed for the procedure \FMS($t,t'$) is at most $2\delta$, where $\delta$ is
the number of vertices in $\Delta$. From the description, it is
obvious that we need only $O(1)$ extra-space.
\end{proof}

%
%
%
%
%
%
%

\vspace{-0.18in}
\paragraph{Time complexity analysis}
The complexity of each do-while loop is the time needed for the procedure \FMS\
which is at most $2\delta$ (see Lemma~\ref{l4}). As the value of the variable 
$Size$ is incremented
by $\delta$ after each iteration of the do-while loop, the time complexity of
our algorithm is $O(2\times Size)$, where $Size$ is the final value of $Size$
after the completion of the do-while loop. As the maximum  value of $Size$
(i.e $|T_{t'}(t)|$) is bounded by $n$, so the time complexity of the algorithm
is $O(n)$. Thus, we have the following:
\vspace{-0.1in}
\begin{theorem}\label{th-centroid}
 The centroid of a tree, given in a read-only memory, can be computed in $O(n)$
time using $O(1)$ extra-space.
\end{theorem}
\vspace{-0.1in}

\vspace{-0.1in}
\section{Weighted 1-center of a tree}\label{w-1-c}
\vspace{-0.1in}

%
%
%
%
%

 Our approach to compute the weighted 1-center of a tree in
constant-work-space model is similar to the
$O(n\log n)$ time $O(n)$ extra-space algorithm proposed by Kariv
and Hakimi~\cite{KH-1}. Overview of the  algorithm is as follows. First, it 
finds an edge $e^*$ where the center of the weighted 1-center  lies. Next,  it
finds the absolute weighted 1-center on that edge $e^*$ using prune-and-search.

\vspace{-0.13in}
\subsection{Finding the edge $e^*$}\label{w-1-c: edge-find}\vspace{-0.09in}
Kariv and Hakimi's~\cite{KH-1}  prune-and-search based algorithm for finding
the edge $e^*$ is based on the following: 
\vspace{-0.08in}
\begin{lemma}\label{w-1-c:l1}
 If $c$ is a fixed vertex of the tree $T=(V,E)$ and $v'$ is a vertex in the
subtree $T_{t}(c)$ ($t\in N(c)$) satisfying
$w(v')d(v',c)=\max_{v\in V}
w(v)d(v,c)$, then the 1-center of $T$ is in the subtree $T_{t}(c^+)$.
\end{lemma}
\vspace{-0.1in}

It initializes  $T'=T$. In 
each iteration, it finds the centroid $c$ of  $T'$, 
and identifies  a vertex $v'$ satisfying Lemma \ref{w-1-c:l1} by traversing 
all the 
vertices of $T$. Thus the subtree ${T}'_t(c)$, containing $v'$, is 
identified. Then, it sets $T'=T'\cap {T}'_t(c^+)$, and unless $T'$ is an edge it
repeats the next iteration.  Since $c$ is the centroid, in each  iteration a
subtree containing at least $\frac{|T'|}{2}$ vertices is pruned.  Thus, the number of iterations is $O(\log n)$ in
the worst case.
 Though the time complexity of computing the 
centroid  is $O(|T'|)$ \cite{KH-1}, the time
taken for
identifying the subtree  of $c$ containing $v'$ is $O(|T|)$ as it needs to
traverse all the vertices of $T$. Thus, the
overall time complexity of this algorithm is $O(n\log n)$. 

In Section \ref{centroid}, we have already shown that centroid of a tree
$T'$ can be computed in $O(|T'|)$ time 
in the constant-work-space model. In order to make this algorithm work in
constant-work-space model, we have to make sure that $T'$ can be
identified from $T$ using $O(1)$ extra-space.
\begin{wrapfigure}{r}{0.5\textwidth} \vspace{-0.2in}    
\begin{minipage}[c]{0.25\textwidth} 
\begin{center} %
\includegraphics[scale=0.25]{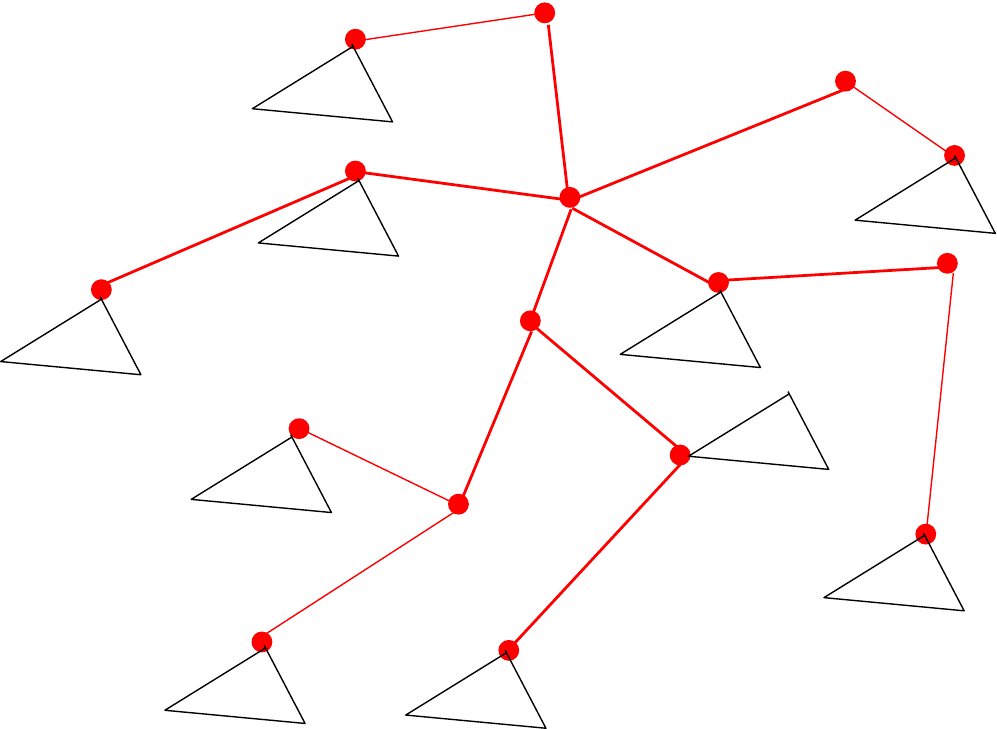}\\
(a)  
\end{center}
\end{minipage}%
\begin{minipage}[c]{0.25\textwidth}
\hspace{-0.1in}
\begin{center} %
\includegraphics[scale=0.25]{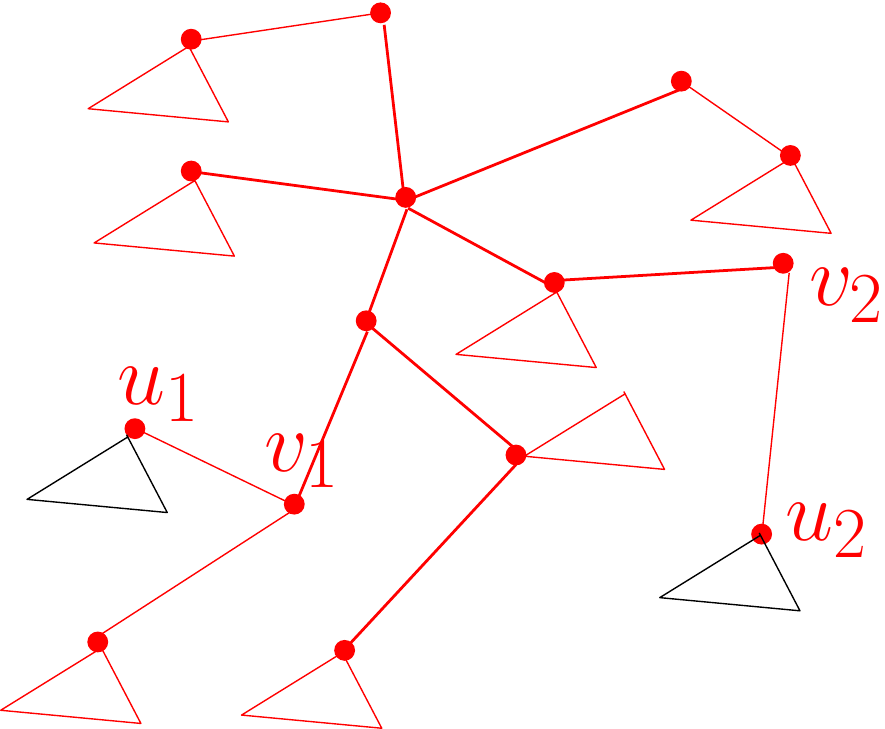}\\
(b)  
\end{center}
\end{minipage}
    \vspace{-0.1in}
\caption{ $T'$ is the red portion of the tree 
} 
\label{fig3}
    \vspace{-0.1in}
\end{wrapfigure}
Observe that after each iteration at most two internal nodes of $T$ may become
leaves of $T'$. So $T'$ may 
have $O(\log n)$ leaves which are internal vertices of the original tree $T$ 
(see 
Figure~\ref{fig3}(a) where red portion indicates $T'$). Such a
representation of $T'$ can not be encoded using $O(1)$ extra-space. To overcome
this, we modify the
algorithm maintaining the following invariant:
%

%

%
%
\vspace{-0.07in}
\begin{invariant}\label{w-1-c:inv}
 At most two internal vertices of $T$ are leaves  of $T'$.
\end{invariant} \vspace{-0.07in}

This invariant enables us to encode $T'$ using only four variables
$u_1,v_1,u_2$ and $v_2$ as follows.
\begin{itemize}
 \item If $(u_i,v_i)\neq(\emptyset,\emptyset)$ for both $i=1$ and $2$ (i.e $T'$ has two internal nodes of $T$ as
leaves), then $T'=T_{v_1}(u_1^+) \cap T_{v_2}(u_2^+)$ ((see 
Figure~\ref{fig3}(b));
\item Else if $(u_i,v_i) \neq (\emptyset,\emptyset)$ for $i=1$ or $i=2$
(i.e $T'$ has one internal
node of $T$ as leaf),  then
$T'=T_{v_i}(u_i^+)$;
\item Else $T'=T$.
\end{itemize} \vspace{-0.09in}
We use another variable $root'$ which signifies the root of the tree $T'$. If
any one of $(u_i,v_i)\neq (\emptyset,\emptyset)$ and $(u_i)$ is the parent of
$v_i$, then $root'=u_i$; otherwise, $root'=root(T)$.

The pseudo-code of our algorithm is given in  Appendix-2 as
Algorithm~\ref{Alg-Weighted1Center}. In each iteration, we compute the centroid
$c$ of $T'$ as stated in Section~\ref{centroid}. Next, by traversing the whole
tree $T$, we find the subtree $T_{t}(c^+)$ which contain  a vertex $v'$
satisfying Lemma~\ref{w-1-c:l1}.
Here one of the following  two situations arises. 
 (i) If at most one of $u_i$ $i\in\{1,2\}$ is in
the subtree $T_{t}(c^+)$, then $T'\cap T_{t}(c^+)$ has at most two internal
nodes of $T$ as leaves. 
(ii) Otherwise, if both $u_1$ and $u_2$  are  in  the subtree $T_{t}(c^+)$, 
then $T'\cap T_{t}(c^+)$ may have at most  three internal nodes of $T$ (namely 
$c$ , $u_ 1$
and $u_2$) as leaves. We can test this in $O(n)$ time using $O(1)$ extra-space.
In the former case, we
 set $T'=T'\cap T_{t}(c^+)$ by updating $(u_i,v_i)$ for the desired
$i\in\{1,2\}$. In the
later case, if three internal vertices appear in   $T'\cap T_{t}(c^+)$, we do
the following.                
 \vspace{-0.1in}                                                               
  
\begin{itemize}
 \item First, we
compute the {\it junction($c,u_1,u_2$)} which is a vertex $j$ of the subtree
$T'\cap T_{t}(c^+)$ such that $c$, $u_1$ and $u_2$ are in three different
subtrees $T_{k_\ell}(j)$, $k_\ell\in N(j)$ for $\ell=1,2,3$. It is left to the
reader to verify that one can compute the
{\it junction($c,u_1,u_2$)} in $O(n)$ time using $O(1)$ extra-space. 

\item Next, we compute $S(j,T)=\max_{v\in V} w(v)d(v,j)$   and find
 an adjacent vertex of $j$ such that the subtree ${T}_{t'}(j)$
contains a vertex $v''$ for which $w(v'')d(v'',j)=\max_{v\in V}
w(v)d(v,j)$ (satisfying Lemma~\ref{w-1-c:l1}).
Note that, as $j$ is the {\it junction($c,u_1,u_2$)}, ${T}_{t'}(j)$ can contain at
most one
of $u_1$, $u_2$ and $c$. As a result, $T'\cap T_{t}(c^+)\cap
{T}_{t'}(j)$ has at most two internal nodes of $T$ as leaves.  So, we
appropriately update $T'=T'\cap T_{t}(c^+)\cap {T}_{t'}(j)$ by updating
$(u_i,v_i)$ for
$i=1,2$.
\end{itemize}
\vspace{-0.1in}

Using induction, we can prove that the
Invariant~\ref{w-1-c:inv} is maintained after each iteration. 
 This is to be
observed 
that $T'$ is decreased by at least half after each iteration. Thus,  we have the following result.

\vspace{-0.05in}
\begin{lemma}
 For a tree $T$ given in a read-only-memory, one can obtain the edge $e^*$ 
where the center of the weighted 1-center lies in $O(n\log n)$ time using $O(1)$
extra-space. 
\end{lemma}

\vspace{-0.18in}
\subsection{Computing weighted 1-center on the  edge $e^*$}

We  find the weighted 1-center $c^*$ on the edge $e^*=(u^*,v^*)$ using
prune-and-search
algorithm similar to Section~\ref{CMEC}. Let $V_1$ and $V_2$ be the set of
vertices in the tree $T_{u^*}(v^*)$  and  $T_{v^*}(u^*)$, respectively. Each
vertex $v\in V_1$ (resp. $v\in V_2$) contributes a linear function $f_1(v,x)= w(v)
d(u^*,v)
+w(v)d(x,u^*)$ (resp. $f_2(v,x)= w(v) d(v^*,v) +w(v)d(x,v^*)$) which signifies
the
weighted distance from $v$ to a point $x\in e^*$. In this regard, it is easy to prove the following:
\vspace{-0.08in}
\begin{lemma}
 All the vertices  $v\in V_1$ (resp. $v\in V_2$) 
and  the corresponding distance $d(u^*,v)$ (resp. $d(v^*,v)$) can be
enumerated in some order in $O(n)$ time using $O(1)$ extra-space.
\end{lemma}
\vspace{-0.1in}
 As in
Section~\ref{CMEC}, given a point $m\in (a,b)$ here also we use {\it Query($m$)}
to decide whether $m$ is the $c^*$ or $c^*$ lies in  $(a,m)$ or in $(m,b)$.  
For each vertex $v\in V_1$ (resp. $v\in V_2$), we compute the $f_1(v,m)$
(resp. $f_2(v,m)$) and find the one with maximum  $f_1(v,m)$ (resp.
$f_2(v,m)$) value. Let $k_1\in V_1$ and $k_2\in V_2$ be two vertices for which the
$f_{1}(k_1,m)$ and $f_{2}(k_2,m)$ are maximum, respectively. If  $f_{1}(k_1,m)
=f_{2}(k_2,m)$, then $m$ is the $c^*$; else  if $f_{1}(k_1,m)
>f_{2}(k_2,m)$, then $c^* \in (a,m)$, otherwise $c^* \in (m,b)$. So, we can
answer {\it Query($m$)} in $O(n)$ time using $O(1)$ space.
 
Here we define the {\em dominance} relation as follows:

\vspace{-0.1in}
\begin{defn}\label{def3}
For a pair of vertices $p,q \in V_1$ (resp.  $p,q \in V_2$ ),
  $p$ is said to {\em dominate} $q$ with respect to a
feasible region $U=(a,b)$, if their corresponding functions $f_1(p,x)$ and
$f_1(q,x)$ (resp. $f_2(p,x)$ and $f_2(q,x)$ ) does not
intersect within the feasible region $U=(a,b)$ and the value of
$f_1(q,x)<f_1(p,x)$ (resp. $f_2(q,x)<f_2(p,x)$ ) for  $x\in U=(a,b)$.
\end{defn}\vspace{-0.1in}
Note that if $f_1(p,x)$ and $f_1(q,x)$ do not intersect  within the feasible 
region $U=(a,b)$, then by checking at any point $x\in U=(a,b)$, we can decide
which one is dominating. It is left to the reader to verify that  this relation
also satisfies the following lemma.
\vspace{-0.1in}
\begin{lemma}
 \label{L10}
 If $p$ dominates $q$ and $q$ dominates $r$ with respect to a feasible region
$U=(a,b)$, then $p$ dominates $r$ with respect to the feasible region
$U=(a,b)$.
\end{lemma}
\vspace{-0.1in}

Now, we follow  the pairing strategy as given in Section~\ref{CMEC}.
Initially, we consider that all the vertices in $V_1$ (resp. $V_2$) are
{\it valid} and the feasible region for $c^*$ is $U=(a,b)=(u^*,v^*)$.
At the beginning of  each phase,  we pair up the consecutive valid elements of $V_1$ (resp. 
$V_2$) in a similar fashion as described in Section~\ref{CMEC}. A  pair of
vertices 
$(v_1,v_2)$  contributes an intersection point
$i(v_1,v_2)=\frac{w(v_1)d(u^*,v_1)-w(v_2)d(u^*,v_2)}{w(v_2)-w(v_1)}$ 
of their corresponding function $f_1$ (resp. $f_2$). A constructed pair
$(v_1,v_2)$ is
considered as a {\it valid pair} with respect to $U$ if the corresponding
intersection point
$i(v_1,v_2)$ lies in $U$, otherwise we can prune one of $v_1$ and $v_2$
depending on whose $f_{1}$ 
(resp. $f_{2}$)  value is less in $(U)$.  We
find the median $m$ of these intersection values considering all the valid pairs and perform
{\it Query($m$)}. So, after this we can prune one element each from  at least
$\frac{1}{4}$-th of the valid  pairs. After at most $O(\log n)$ iterations, we
can prune one element from each of the valid pairs. 
%
%
%
%
%
%
%
%
%
Following the same frame-work as given in Section~\ref{CMEC}, we have the
following:
\vspace{-0.15in}
\begin{theorem}\label{th-w-1-c}
 The weighted 1-center of a tree $T$ can be computed in $O((n+M)\log^2 n)$ time
using $O(1)$ extra-space in the constant-work-space model, where $M$ is the time
needed to compute the median of
$n$ elements given in a read-only memory when $O(1)$ space is
provided.
\end{theorem}
\vspace{-0.15in}
\begin{remark}
 We can compute the weighted median and weighted 2-center of a tree in $O(n)$
and   $O((n+M)\log^2 n)$ time, respectively, where $M$ is the time
needed to compute the median of
$n$ elements given in a read-only memory when $O(1)$ space is
provided. The detail is given in the Appendix.
\end{remark}
\vspace{-0.18in}
\section{Concluding Remarks}\vspace{-0.12in}
In this paper, we present some fundamental facility location problems in
constant-work-space model. The selection problem plays a crucial role in the
complexity of the algorithms. Randomized selection could be used to make the
algorithms faster.
The strategy to compute prune-and-search using constant-space can be used to
solve two and three dimensional linear programming in $O(n~polylog(n))$ time and
$O(1)$ extra-space.
 We believe that some of the techniques used here
can be helpful to solve other relevant  problems as well. It would be worthy to
study similar problems in general graphs such as cycle,
monocycle, cactus etc. in the constant-work-space model.  
\vspace{-0.18in}
\small
\bibliographystyle{abbrv}
\bibliography{references}

\begin{thebibliography}{10}

\bibitem{AgarwalS10}
P.~K. Agarwal and R.~Sharathkumar.
\newblock Streaming algorithms for extent problems in high dimensions.
\newblock In {\em SODA}, pages 1481--1489, 2010.

\bibitem{AllenderM04}
E.~Allender and M.~Mahajan.
\newblock The complexity of planarity testing.
\newblock {\em Inf. Comput.}, 189(1):117--134, 2004.

\bibitem{AroraB}
S.~Arora and B.~Barak.
\newblock {\em Computational Complexity - A Modern Approach}.
\newblock Cambridge University Press, 2009.

\bibitem{AsanoBBKMRS13}
T.~Asano, K.~Buchin, M.~Buchin, M.~Korman, W.~Mulzer, G.~Rote, and A.~Schulz.
\newblock Memory-constrained algorithms for simple polygons.
\newblock {\em Comput. Geom.}, 46(8):959--969, 2013.

\bibitem{AsanoMRW11}
T.~Asano, W.~Mulzer, G.~Rote, and Y.~Wang.
\newblock Constant-work-space algorithms for geometric problems.
\newblock {\em JoCG}, 2(1):46--68, 2011.

\bibitem{AsanoMW11}
T.~Asano, W.~Mulzer, and Y.~Wang.
\newblock Constant-work-space algorithms for shortest paths in trees and simple
  polygons.
\newblock {\em J. Graph Algorithms Appl.}, 15(5):569--586, 2011.

\bibitem{Ben-MosheBS06}
B.~Ben-Moshe, B.~K. Bhattacharya, and Q.~Shi.
\newblock An optimal algorithm for the continuous/discrete weighted 2-center
  problem in trees.
\newblock In {\em LATIN}, pages 166--177, 2006.

\bibitem{Chan10}
T.~M. Chan.
\newblock Comparison-based time-space lower bounds for selection.
\newblock {\em ACM Transactions on Algorithms}, 6(2), 2010.

\bibitem{ChanC07}
T.~M. Chan and E.~Y. Chen.
\newblock Multi-pass geometric algorithms.
\newblock {\em Discrete {\&} Computational Geometry}, 37(1):79--102, 2007.

\bibitem{ChanP11}
T.~M. Chan and V.~Pathak.
\newblock Streaming and dynamic algorithms for minimum enclosing balls in high
  dimensions.
\newblock In {\em WADS}, pages 195--206, 2011.

\bibitem{De-Minati-Thesis}
M.~De.
\newblock {\em Space-efficient Algorithms for Geometric Optimization Problems}.
\newblock PhD thesis, Indian Statistical Institute, 2013.

\bibitem{DeNR12b}
M.~De, S.~C. Nandy, and S.~Roy.
\newblock Minimum enclosing circle with few extra variables.
\newblock In {\em FSTTCS}, pages 510--521, 2012.

\bibitem{Hakimi65}
S.~L. Hakimi.
\newblock Optimum distribution of switching centers in a communication network
  and some related graph theoretic problems.
\newblock {\em Operations Research}, 13(3):462--475, 1965.

\bibitem{Harary}
F.~Harary.
\newblock {\em Graph Theory}.
\newblock Addison-Wesley, 1972.

\bibitem{KH-1}
O.~Kariv and S.~L. Hakimi.
\newblock An algorithmic approach to network location problems. i: The
  p-centers.
\newblock {\em SIAM Journal on Applied Mathematics}, 37(3):513--538, 1979.

\bibitem{KH-2}
O.~Kariv and S.~L. Hakimi.
\newblock An algorithmic approach to network location problems. ii: The
  p-medians.
\newblock {\em SIAM Journal on Applied Mathematics}, 37(3):539--560, 1979.

\bibitem{Megiddo83a}
N.~Megiddo.
\newblock Linear-time algorithms for linear programming in ${\IR}^3$ and
  related problems.
\newblock {\em SIAM J. Comput.}, 12(4):759--776, 1983.

\bibitem{MunroP78}
J.~I. Munro and M.~Paterson.
\newblock Selection and sorting with limited storage.
\newblock In {\em FOCS}, pages 253--258, 1978.

\bibitem{MunroR96}
J.~I. Munro and V.~Raman.
\newblock Selection from read-only memory and sorting with minimum data
  movement.
\newblock {\em Theor. Comput. Sci.}, 165(2):311--323, 1996.

\bibitem{RamanR99}
V.~Raman and S.~Ramnath.
\newblock Improved upper bounds for time-space trade-offs for selection.
\newblock {\em Nord. J. Comput.}, 6(2):162--180, 1999.

\bibitem{Reingold05}
O.~Reingold.
\newblock Undirected st-connectivity in log-space.
\newblock In {\em STOC}, pages 376--385, 2005.

\bibitem{Qiaosheng-Shi-Thesis}
Q.~Shi.
\newblock {\em Efficient algorithms for network center/covering location
  optimization problems}.
\newblock PhD thesis, Simon Fraser University, 2008.

\end{thebibliography}
\normalsize

%
%
%
%
%
%
%
\newpage
\section{Appendix:}

\subsection{Appendix-1: Weighted median}\label{WeightedM}
%
%

%

Based on the fact  that  a vertex
$v$ of a tree ${T}$ is  {\it weighted-centroid} if
and only if $v$ is {\it weighted median}~\cite{KH-2}, we present
an $O(n)$ time algorithm to find the weighted median of a tree
using $O(1)$ extra-space. The pseudo-code of the algorithm is given
in Algorithm~\ref{Alg-WeightedMedian}.  The structure of the 
Algorithm~\ref{Alg-WeightedMedian} is similar to the
Algorithm~\ref{Alg-Centroid}.

\begin{algorithm}[b]
\scriptsize
\SetAlgoLined
\caption{\WeightedMedian($T$)}
\label{Alg-WeightedMedian}
\KwIn{A tree $T$ is given in a read-only memory}
\KwOut{Report the centroid of the given tree}
\SetKwRepeat{Repeat}{do}{while}

$t=root(T)$; $t'=\emptyset$; $Size=0$; $WSize=0$\;
$SumWeight=w(T)=\sum_{v\in V}w(v)$ ; /* Can be computed by traversing the
tree*/ \\

\Repeat{$WTMSize \leq\lfloor\frac{SumWeight}{2}\rfloor$}
{
  $(m,\delta,\delta_w,WTMSize)=\FMWS(t,t',Size,WSize)$\;
  \If{$WTMSize > \lfloor\frac{SumWeight}{2}\rfloor$}
  {$t'=t$; $t= m$;  $WSize= WSize + \delta_w$; $Size= Size +\delta$  \;}
}

Return $t$\;
\normalsize
\end{algorithm}

First, the algorithm  computes  $w(T)=\sum_{v\in V}w(v)$ and keeps it
in the  variable $SumWeight$.
Note that this can be computed by traversing the whole tree in $O(n)$ time
using $O(1)$ extra-space. 
As in the Section~\ref{centroid}, here also the variables  $t$, $t'$, $Size$
and $WSize$ maintain the following invariant:
\vspace{-0.09in}
\noindent
\begin{invariant}
\begin{itemize}
\item[] Initially,  $t=root(T)$, $t'=\emptyset$,  $Size=0$ and $WSize=0$.
\vspace{-0.05in}
\item[] If $t'\neq \emptyset$, then 
$t$ and $t'$ are adjacent vertices of $T$, $Size=|T_{t'}(t)|$ and
$WSize=w(T_{t'}(t))$
\end{itemize}
\end{invariant}
\vspace{-0.05in}


At each iteration of the do-while loop,  the  algorithm  evokes the  procedure 
\FMWS$(t,t',Size,WSize)$
which returns four parameters $m$, $\delta$, $\delta_w$ and $WTMSize$. 
Here  $m$ is the adjacent vertex of $t$ such that
$w(T_m(t))=MaxWS(t)=\max_{v'\in
N(v)}w(T_{v'}(t))$, $WTMSize=w(T_{m}(t))$, $\delta$ is the number of vertices
in
the subtree $\Delta$ and $\delta_w=w(\Delta)$, where $\Delta=
T\setminus\{T_{t'}(t)\cup
T_{m}(t) \}$ (see Figure~\ref{fig2}).
If $WTMSize\leq\lfloor\frac{SumWeight}{2}\rfloor$, then the algorithm
terminates with reporting $t$ as the weighted median; otherwise it updates $t$
and $t'$ by setting $t'=t$ and $t=m$, and repeats the while-loop.
The correctness of
this algorithm follows from the fact that $v$ is a weighted-centroid of a tree
if and only if $MaxWS(v)\leq \frac{SumWeight}{2}$~\cite{KH-2}.  
Similar to the Lemma~\ref{l4}, we can prove the following lemma.

\begin{lemma}\label{l5}
 The procedure \FMWS($t,t', Size, WSize$), which returns four parameters $m$,
$\delta$, $\delta_w$
and $WTMSize$,  can be implemented in $O(2\delta)$  time using $O(1)$
extra-space. Here  $m$
is the adjacent vertex of $t$ such that $w(T_m(t))=\max_{v'\in
N(v}w(T_{v'}(t))$, $WTMSize=w(T_{m}(t))$, $\delta$ is the number of vertices
in
the subtree $\Delta$ and $\delta_w=w(\Delta)$, where $\Delta=
T\setminus\{T_{t'}(t)\cup
T_{m}(t) \}$.
\end{lemma}\vspace{-0.06in}

\begin{proof}
The main difference with the procedure \FMS\ is that  the procedure
\FMWS\ computes the maximum weighted subtree instead of maximum sized subtree. 
Note that, using the three routines
$Parent(v')$, $FirstChild(v')$, 
$NextChild(t,v')$,  one can compute the total weight of all the vertices  in the
subtree $T_{v'}(t)$ by a depth-first traversal. It takes $O(|T_{v'}(t)|)$
 time and  $O(1)$ extra-space.  Thus we can implement the procedure \FMWS($t$,
$t', Size,WSize$)
 similar to the procedure \FMS\  in $O(2\delta)$
time using $O(1)$ extra-space (see Lemma~\ref{l4}). 
\end{proof}

%
%


For the similar reason given while analyzing the time complexity of the
Theorem~\ref{th-centroid}, we can argue that the time complexity of the  
Algorithm~\ref{Alg-WeightedMedian} is $O(n)$.
Thus, we have the following:

\begin{theorem}
 The weighted median of a tree, given in a read-only memory, can be computed in
$O(n)$ time using $O(1)$ extra-space.
\end{theorem}
\vspace{-0.18in}



\subsection{Appendix-2}

 \begin{algorithm}[!h]
\SetAlgoLined
\caption{\EdgeFindOC($T$)}
\label{Alg-Weighted1Center}
\KwIn{A tree $T$ is given in a read-only memory}
\KwOut{Report the edge $e^*$ on which weighted 1-center lies}

 $(u_1,v_1)=(\emptyset,\emptyset)$;  $(u_2,v_2)=(\emptyset,\emptyset)$;\
/*  $T'=T$  */\\
 \While{$T'$ is not an edge}
 {
 Find the centroid $c$ of the tree $T'$.\\
 Let  $t$ be an adjacent vertex of $c$ such that the subtree ${T}'_{t}(c)$
contains a vertex $f$ for which 
$w(f)d(f,c)=\max_{v\in V} w(v)d(v,c)$.\\
 \If{$(u_1,v_1)\neq(\emptyset,\emptyset)\wedge 
(u_2,v_2)\neq(\emptyset,\emptyset)$ and $T_{t}(c^+)$ contains both $u_1$ and
$u_2$}
{
$j=$ Junction$(c,u_1,u_2)$\;

 Let  $t'$ be an adjacent vertex of $j$ such that the subtree ${T}_{t'}(j)$
contains a vertex $f'$ for which 
$w(f')d(f',j)=\max_{v\in V} w(v)d(v,j)$.\\
\If{$T_{t'}(j)$ contains $(u_i,v_i)$ where $i=1$ or $2$}
{$(u_{3-i},v_{3-i})=(j,t')$; /*  $T'=T'\cap {T}_{t}(c^+) \cap T_{t'}(j)$ */\\ }
\ElseIf{$T_{t'}(j)$ contains $c$}{$(u_1,v_1)=(j,t')$;
$(u_2,v_2)=(c,t)$; /*  $T'=T'\cap {T}_{t}(c^+) \cap T_{t'}(j)$ */\\}
\Else{$(u_1,v_1)=(j,t')$; $(u_2,v_2)=(\emptyset,\emptyset)$; /*  $T'=T'\cap
{T}_{t}(c^+)\cap T_{t'}(j)$ */\\}
 
 }
 \ElseIf{$(u_i,v_i)\neq(\emptyset,\emptyset)$  
 and $T_{t}(c^+)$ contains  $(u_i,v_i)$ where $i=1$ or
$2$}{$(u_{3-i},v_{3-i})=(c,t)$; /*  $T'=T'\cap {T}_{t}(c^+)$ */\\}

  \Else{$(u_1,v_1)=(c,t)$; $(u_2,v_2)=(\emptyset,\emptyset)$; /*  $T'=T'\cap
{T}_{t}(c^+)$ */\\}

 }
 Report $T'$\;
\end{algorithm}

\newpage
\subsection{Appendix-3: Weighted 2-center}


%
%
%
%
We can obtain the weighted 2-center of a tree $T$ in the constant-work-space
model  based on  the  $O(n\log n)$ time
algorithm proposed by Ben-Moshe et al.~\cite{Ben-MosheBS06}.
The overview of the algorithm is follows. First, it finds the {\it split edge}
$e^*=(u^*,v^*)$ of  $T$ which satisfies the following:

$\max\{r(T_{u^*}(v^*)),
r(T_{v^*}(u^*))\}=\min_{e:(u,v)\in E}max\{r(T_{u}(v)),r(T_{v}(u))\}$,\\
 where
$r(T)=\min_{x\in T}\max_{v\in V}w(v)d(x,v)$ is the weighted radius of the tree $T$. 
The split edge partitions the tree into two parts $T_1=T_{u^*}(v^*)$  and $T_2=T_{v^*}(u^*)$. Finally,
the algorithm finds the weighted 1-center of $T_1$ and $T_2$ separately. These
two weighted 1-centres are actually the weighted 2-center of the whole tree
$T$. In the previous section, we have already presented an algorithm to compute weighted 1-center in
a tree in the constant-work-space model. Thus we only have to
show that we can find
the split edge in this model.

%
%


\subsubsection{Finding the optimal split edge $e^*$}

 Note that we can compute
$r(T_{t}(v))$ in constant-work-space model by the following way. First, we
compute the weighted 1-center $c_t$ of $T_{t}(v)$ by Theorem~\ref{th-w-1-c}.
Next, we find 
the maximum distance from any vertex of $T_{t}(v)$ to $c_t$ by
traversing the tree $T_{t}(v)$. This distance is the $r(T_{t}(v))$.
Given a vertex $v$ in $T$ we can decide the  subtree $T_{t}(v)$, $t\in
N(v)$ in which the optimal split edge $e^*$ lies by evaluating $r(T_{t}(v))$ for
all $t\in
N(v)$ (see Lemma 4 in \cite{Ben-MosheBS06}). So, we
can make this decision in our constant-work-space model in linear time. 

The algorithm for finding the split edge $e^*$   works almost in a similar way as
we have computed the
edge on which weighted 1-center lies in Section~\ref{w-1-c: edge-find}. The
pseudo-code is given in Algorithm~\ref{Alg-SplitEdge}. 
Thus, we have the following result.

\begin{theorem}
 Weighted 2-center of a tree $T$ can be computed in in $O((n+M)\log^2 n)$ time
using $O(1)$ extra-space in the constant-work-space model, where $M$ is the time
needed to compute the median of
$n$ elements given in a read-only memory when $O(1)$ space is provided.
\end{theorem}

 \begin{algorithm}[!h]
\scriptsize
\SetAlgoLined
\caption{\FindSEdge($T$)}
\label{Alg-SplitEdge}
\KwIn{A tree $T$ is given in a read-only memory}
\KwOut{Report the split edge $e^*$ }

 $(u_1,v_1)=(\emptyset,\emptyset)$;  $(u_2,v_2)=(\emptyset,\emptyset)$;\
/*  $T'=T$  */\\
 \While{$T'$ is not an edge}
 {
 Find the centroid $c$ of the tree $T'$.\\
 \blue{Let  $t$ be an adjacent vertex of $c$ such that the subtree ${T}'_{t}(c)$
contains the optimal split edge.}\\
 \If{$(u_1,v_1)\neq(\emptyset,\emptyset)\wedge 
(u_2,v_2)\neq(\emptyset,\emptyset)$ and $T_{t}(c^+)$ contains both $u_1$ and
$u_2$}
{
$j=$ Junction$(c,u_1,u_2)$\;

 \blue{Let  $t'$ be an adjacent vertex of $j$ such that the subtree
${T}_{t'}(j)$
contains the optimal split edge .}\\
\If{$T_{t'}(j)$ contains $(u_i,v_i)$ where $i=1$ or $2$}
{$(u_{3-i},v_{3-i})=(j,t')$/*  $T'=T'\cap {T}_{t}(c^+) \cap T_{t'}(j)$ */\\}
\ElseIf{$T_{t'}(j)$ contains $c$}{$(u_1,v_1)=(j,t')$;
$(u_2,v_2)=(c,t)$/*  $T'=T'\cap {T}_{t}(c^+) \cap T_{t'}(j)$ */\\}
\Else{$(u_1,v_1)=(j,t')$; $(u_2,v_2)=(\emptyset,\emptyset)$/*  $T'=T'\cap
{T}_{t}(c^+) \cap T_{t'}(j)$ */\\}
 
 }
 \ElseIf{$(u_i,v_i)\neq(\emptyset,\emptyset)$  
 and $T_{t}(c^+)$ contains  $(u_i,v_i)$ where $i=1$ or
$2$}{$(u_{3-i},v_{3-i})=(c,t)$; /*  $T'=T'\cap {T}_{t}(c^+)$ */\\}

  \Else{$(u_1,v_1)=(c,t)$; $(u_2,v_2)=(\emptyset,\emptyset)$; /*  $T'=T'\cap
{T}_{t}(c^+)$ */\\}

 }
 Report $T'$
 \normalsize
\end{algorithm}

\end{document}